\documentclass{JHEP3}

\usepackage{amsmath}
\usepackage{amsfonts}
\usepackage{amssymb}
\usepackage{graphicx}
\usepackage{epsfig}
\usepackage{bbm}

\def\bfx{{\mathbf x}}
\def\bfy{{\mathbf y}}

\def\plinks{\mathcal{L}^{+}_{\rightarrow}}

\newcommand{\tr}{\mathrm{tr} \ }
\newcommand{\Tr}{\mathrm{tr} \ }
\newcommand{\comma}{\ ,}
\newcommand{\period}{\ .}
\newcommand{\Real}{\mathrm{Re}}
\newcommand{\Imag}{\mathrm{Im}}
\newcommand{\R}{R}
\newcommand{\Z}{Z}

\title{Center symmetry and the orientifold planar equivalence
}

\author{Luigi Del Debbio \\ 
        SUPA, School of Physics and Astronomy, University of Edinburgh \\
	Edinburgh EH9 3JZ, Scotland\\
	E-mail: \email{luigi.del.debbio@ed.ac.uk} 
}
\author{Agostino Patella \\
        Department of Physics, University of Swansea\\
	Swansea SA2 8PP, Wales\\
        E-mail: \email{a.patella@swan.ac.uk} 
}

\abstract{ We study the center symmetry of $\mathrm{SU}(N)$ gauge
theories with fermions in the two--index representations, by computing
the effective potential of the Polyakov loop in the large--mass
expansion on the lattice. In the large--$N$ limit and at non--zero
temperature, we find that the center symmetry is $\Z_N$ for fermions
in the adjoint representation and just $\Z_2$ for fermions in the
(anti)symmetric representation. We discuss the fact that our results
do not contradict the orientifold planar equivalence, which relates a
common sector defined by the bosonic gauge--invariant $C$--even states
of theories with fermions in different two--index representations. Our
results complement the work of Armoni \textit{et al.} (2007), who
showed how at zero temperature a $\Z_N$ center symmetry is dynamically
recovered also for fermions in the (anti)symmetric representation, by
considering the theories at finite temperature. }


\preprint{}

\begin{document}

\section{Introduction}
\label{ref:intro}

Confinement of colour is related to center symmetry: where center
symmetry is present and not broken, it implies confinement. Its
spontaneous breaking leads to the deconfinement transition, and the
critical behaviour at the phase transition is described by a
three--dimensional effective theory which encodes the relevant pattern
of symmetry breaking~\cite{Svetitsky:1982gs}.

On the other hand, orientifold planar equivalence is the
equivalence in the large--$N$ limit of the $\mathrm{SU}(N)$ gauge theory with a
Dirac fermion in the symmetric/antisymmetric two--index
representations (OrientiQCD) and the $\mathrm{SU}(N)$ gauge theory with a
Majorana fermion in the adjoint representation (AdjQCD) in a common
sector~\cite{Armoni:2003fb,Armoni:2003gp,Armoni:2004ub}. The validity
of this equivalence was discussed in detail in
Refs.~\cite{Unsal:2006pj,Armoni:2007rf}. A proof of the equivalence on
the lattice in the strong--coupling and large--mass phase can be found
in Ref.~\cite{Patella:2005vx} (a general setup for descussing planar
equivalences between theories with two-index representations in the
strong--coupling and large--mass phase was also presented
in~\cite{Kovtun:2003hr,Kovtun:2004bz}).

A potential inconsistency was pointed out in
Ref.~\cite{Sannino:2004qp}: the two theories involved in the
equivalence have different symmetries and in particular different
center symmetries. The action of the adjoint theory is invariant under
the full center $\Z_N$ of the gauge group at every value of $N$.
In the orientifold theories the center symmetry is explicitly broken by the
matter to a $\Z_2$ if $N$ is even and to nothing if $N$ is odd.
The different symmetry pattern in
the two theories suggests a different dynamics at the deconfinement
phase transition.  It is therefore interesting to investigate to which
extent the orientifold planar equivalence can describe correctly the
dynamics at the deconfinement phase transition.

A solution to this puzzle in the case of zero temperature was presented in
Ref.~\cite{Armoni:2007kd}: in the orientifold theories the action can
be separated in a part that is invariant under $\Z_N$ and a part that
is not. In the large--$N$ limit, the non--invariant part decouples from
the expectation values of the Wilson loops. The $\Z_N$ symmetry is
therefore dynamically recovered and the apparent contradiction is
resolved. One of the physical consequences is the stability of the
$k$--strings.

In this work we consider the orientifold theories at finite temperature, both in
the confining and deconfining phases. The main difference with the zero
temperature case is that now the theory contains also loops wrapping around the
thermal direction. We show by explicit computations that the non--invariant part
of the action decouples only from the expectation values of the loops with zero
winding number around the thermal direction. In the general case of wrapped
loops, the non--invariant part of the action does not decouple, thus the $\Z_N$
symmetry is not restored in all the sectors of the theory. In particular, we
will show that matter in the orientifold theories explicitly breaks the center
symmetry to the $\Z_2$ subgroup in the large--$N$ limit at finite temperature.

In the deconfined phase, this result is well known. It can be obtained
in the high--temperature regime, by analysing the one--loop effective
action for the Polyakov loop. The invariance under $\Z_2$ but not $\Z_N$
implies that in the deconfined phase (where the $\Z_2$ is spontaneously broken),
only two degenerate vacua exist.

In the large--$N$ limit, the gauge theories discretised on the lattice
exhibit two confined phases as the 't~Hooft coupling, $\lambda=g^2 N$,
is varied: an unphysical strong--coupling phase and a physical
weak--coupling one~\cite{Gross:1980he,Kiskis:2003rd}. In both these
phases, we investigate the symmetry content by computing the
functional generator ($W$) of the connected expectation values of the
Polyakov loop in the framework of the large--mass expansion.  The
generator $W$ is a legitimate tool to study the center symmetry, since
it encodes the same amount of information as the effective potential
of the Polyakov loop. For finite $N$, $W$ can be expanded in powers of
$1/\lambda$:
\begin{equation}
W(\lambda,N)=\sum_n a_n(N) \lambda^{-n} \period
\end{equation}
At strong coupling the large--$N$ limit of $W$ is obtained by taking
the limit of each coefficient $a_n(N)$ in the above
expansion. Properties of Polyakov loops correlators can be inferred
from $W$. In particular we will show that:
\begin{equation}
\lim_{N \rightarrow \infty}
\langle \left( \tr \Omega \right)^2 \rangle_c \neq 0 \comma
\end{equation}
in the orientifold theories, where $(\tr \Omega)$ is the Polyakov
loop; the above result shows clearly that the full $\Z_N$ center is
not a symmetry. We may add that it is explicitly broken (and not spontaneously),
since the vacuum is not degenerate in the strong-coupling regime. Under some
mild assumptions, this result remains valid in the physical (weak--coupling)
confined phase.

Let us now anticipate some comments on the original question. How can
the orientifold planar equivalence be valid if the two theories that
should be equivalent have got different symmetries? It was already
pointed out that the equivalence holds only in a common
\textit{neutral} sector of the various theories. The neutral sector
was identified in Ref.~\cite{Unsal:2006pj} to be the set of all the
single--trace gauge--invariant observables that are also invariant
under charge--conjugation (C) symmetry. The Polyakov loop does not
belong to the neutral sector; indeed under C--symmetry $(\tr \Omega)
\rightarrow (\tr \Omega)^\dagger$, therefore the real part is C--even
and belongs to neutral sector, while the imaginary part is C--odd (it
belongs to the \textit{twisted} sector). Under the action of an
element $e^{\frac{2 \pi i k}{N}}$ of $\Z_N$, the neutral and twisted
sectors are in general mixed:
\begin{subequations}
\begin{flalign}
& \Real \left[\tr \Omega \right] \rightarrow
\cos \left( \frac{2\pi k}{N} \right) \Real \left[\tr \Omega\right] -
\sin \left( \frac{2\pi k}{N} \right) \Imag \left[\tr \Omega \right]\comma \\
& \Imag \left[\tr \Omega \right]\rightarrow
\sin \left( \frac{2\pi k}{N} \right) \Real \left[\tr \Omega \right]+
\cos \left( \frac{2\pi k}{N} \right) \Imag \left[\tr \Omega \right]\period
\end{flalign} \label{eq:intro:center_on_polyakov}
\end{subequations}

The adjoint theory has a $\Z_N$ symmetry, but the neutral and twisted
sectors are mixed by the center symmetry. Only the $\Z_2$ subgroup
(which corresponds to $e^{\frac{2 \pi i k}{N}}=-1$ in
Eqs.~\eqref{eq:intro:center_on_polyakov}) maps the neutral sector into
itself. Therefore the orientifold planar equivalence requires just a
$\Z_2$ symmetry in the neutral sector of the orientifold theories. The
fact that $\Z_2$ (and not $\Z_N$) is the center symmetry for the full
orientifold theories does not follow from the equivalence and is not
in contradiction with it.

The paper is organised as follows. In Sect.~\ref{sec:center} we give a
brief review of center symmetry. In Sect.~\ref{sec:potentials},
\ref{sec:connected}, and~\ref{sec:effpotential} both the
strong--coupling and weak--coupling confining phases on the lattice
are analysed. Although the general setting is valid for both
phases, explicit analytical computations can be performed only in the
strong--coupling phase. In particular, the effective potential in the
large--mass phase is computed. Even though this phase does not
describe continuum physics it is interesting because it provides a
testing bed where the planar equivalence holds and computations can be
performed explicitly.

The possibility of extrapolating some of the results from the
strong--coupling regime to the continuum limit is discussed in
Sect.~\ref{sec:connected}, together with the necessary underlying
assumptions. A brief analysis of the center symmetry in the
high--temperature regime is presented in Sect.~\ref{sec:hitemp}.

Finally in Sect.~\ref{sec:hamiltonian}, we introduce the effective
potential in the Hamiltonian formalism and investigate the center
symmetry by means of the formalism of the coherent
states~\cite{Yaffe:1981vf,Unsal:2006pj}.

\section{Center symmetry}
\label{sec:center}

Whenever a gauge group has non--trivial center, the center symmetry
plays a central role in understanding the dynamics at the
deconfinement phase transition. We briefly summarise in this section
some relevant properties of gauge theories under transformations that
belong to the center of the gauge group. The same notation is used in
the rest of the paper. For a gauge theory discretised on the lattice,
a transformation belonging to the center of the gauge group acts on
the link variables according to:
\begin{equation}
U_0(x_0=0,\mathbf{x}) \rightarrow u U_0(x_0=0,\mathbf{x})\comma
\end{equation}
where $x_0$ is the coordinate along the time direction, and $u$ is an
element of the center. All the other link variables are left
unchanged by the center transformation. The center transformation
counts the winding number of Wilson loops around the time
direction. If the closed path $\Gamma$ wraps $w$ times around the time
direction, the transformation rule is:
\begin{equation}
\tr U(\Gamma) \rightarrow u^w \tr U(\Gamma) \comma
\end{equation}
where $U(\Gamma)$ is the path--ordered product of the link variables
along $\Gamma$. Since the gauge action is given by a sum of
plaquettes, which have zero winding number, it is invariant under the
action of the center.

A generic representation $\R$ of the gauge group induces a
representation of the center subgroup, that can be labelled by an
integer $N_{\R}$ (with $0 \le N_{\R} < N$), the $N$--ality
of the representation:
\begin{equation}
\R[u] = u^{N_\R} \comma \quad \R[uU] = u^{N_\R} \R[U] \period
\end{equation}
Clearly, if matter in a representation with zero $N$--ality is
present, the theory is still invariant under the action of the
center. Otherwise, the theory is invariant only under some
subgroup. The center symmetry of the theories investigated in this
work are summarised in the following table.

\begin{center}
\begin{tabular}{|c|c|c|}
\hline
gauge group & matter & center symmetry \\
\hline
$\mathrm{SO}(2N)$ & none/adjoint & $\Z_{2}$ \\
$\mathrm{SU}(N)$ & none/adjoint & $\Z_{N}$ \\
$\mathrm{SU}(N)$ & fundamental & - \\
$\mathrm{SU}(N)$ with $N$ even & AS/S & $\Z_{2}$ \\
$\mathrm{SU}(N)$ with $N$ odd & AS/S & - \\
$\mathrm{U}(N)$ & none/adjoint & $\mathrm{U}(1)$ \\
$\mathrm{U}(N)$ & fundamental & - \\
$\mathrm{U}(N)$ & AS/S & $\Z_{2}$ \\
\hline
\end{tabular}
\end{center}

\section{Potentials for the Polyakov loop}
\label{sec:potentials}

In this section, we introduce effective potentials that describe the
dynamics of Polyakov loops for the gauge theory discretised on the
lattice. We shall first discuss the case of a pure gauge theory, while
fermions in arbitrary representations are introduced in the next
section within the large--mass expansion framework. In what follows,
it is useful to distinguish the temporal coordinate $x_0$ from the
spatial ones, i.e. $x=(x_0,\mathbf{x})$, and the temporal link
variable $U_0(x_0,\mathbf{x})$ from the spatial ones
$U_{k}(x_0,\mathbf{x})$.

The parallel transport $\Omega$ around the temporal dimension and the
Polyakov loop $P$ are defined as usual by:
\begin{gather}
\Omega(\mathbf{x}) = U_0( 0, \mathbf{x} ) U_0( 1, \mathbf{x} ) \cdots U_0( T-1,
\mathbf{x} ) \comma \\
P( \mathbf{x} ) = 
\tr \Omega(\mathbf{x} ) \comma
\end{gather}
where $T$ is the extent of the lattice in the temporal direction in
units of the lattice spacing.  Gauge invariance is used to fix the
temporal gauge, where we have:
\begin{equation}
U_0(x_0,\mathbf{x}) =
\begin{cases}
1 & \quad \textrm{if } x_0 = 0,\dots,T-2 \\
\Omega(\mathbf{x}) & \quad \textrm{if } x_0 = T-1
\end{cases}
\period
\end{equation}
After such a gauge fixing, the gauge action $S(\Omega,U_{k})$ is
invariant under time--independent gauge transformations. The
Fadeev--Popov determinant being trivial, the measure of the functional
integral is simply:
\begin{equation}
\exp \left\{ -S(\Omega, U_{k}) \right\} \,
\prod_{k,x_0,\mathbf{x}} dU_{k}(x_0,\mathbf{x})
\prod_{\mathbf{x}} d\Omega(\mathbf{x}) \period
\end{equation}

Let us now define several functionals of the field
$\Omega(\mathbf{x})$; they all encode information about the dynamics
of the Polyakov loops, and provide useful information on the structure
of the deconfinement phase transition.

The probability distribution of the parallel transport
$\Omega(\mathbf{x})$ is obtained by integrating out all the link
variables in the spatial directions:
\begin{equation}
e^{ -S_\Omega(\Omega) } = \frac{1}{Z} \int e^{ -S(\Omega, U_k) } \,
\prod_{k,x_0,\mathbf{x}} dU_k(x_0,\mathbf{x})\period
\end{equation}
The functional $S_\Omega(\Omega)$ defines an effective action for the
Polyakov loop, and the vacuum expectation value of a generic
functional $f(\Omega)$ can be computed as:
\begin{equation}
\langle f(\Omega) \rangle = \frac{1}{Z} \int f(\Omega)\,  e^{ -S_\Omega(\Omega) } \,
\prod_\mathbf{x} d \Omega(\mathbf{x}) \period
\end{equation}

The functional $S_\Omega(\Omega)$ induces a natural definition for the
probability distribution of the Polyakov loop as the expectation
value of the delta function:
\begin{eqnarray}
  e^{ -N^2 V(P) } &=& \langle \prod_\mathbf{x} \delta \left( P(\mathbf{x}) -
    \tr \Omega(\mathbf{x}) \right) \rangle = \nonumber \\
  &=& 
  \frac{1}{Z}\, \int d \Omega(\mathbf{x}) \prod_\mathbf{x} \delta \left( P(\mathbf{x}) -
    \tr \Omega(\mathbf{x}) \right) 
  \, e^{ -S_\Omega(\Omega) } \period
\end{eqnarray}
The function $V(P)$ is sometimes referred to as the effective
potential for the Polyakov loop.

In this work, we shall concentrate instead on the quantum action for
the Polyakov loop, which is defined as the Legendre transform of the
free energy of the system. First we construct the generator of connected
$n$--point functions by coupling the Polyakov loop to a complex
external source:
\begin{equation}
  \label{eq:Walpha}
  e^{-N^2 W(\alpha,\bar{\alpha})} = \int d \Omega(\mathbf{x})\, e^{ -S_\Omega(\Omega) } \,
  \exp \left\{ -N \sum_\mathbf{x}\left[\bar{\alpha}(\bfx) P(\bfx) + \alpha(\bfx) 
        P^\dagger(\bfx)\right] \right\} \period
\end{equation}
A Taylor expansion of the exponential in Eq.~(\ref{eq:Walpha}) yields:
\begin{eqnarray}
  W(\alpha,\bar{\alpha}) 
   &=& 1 - \sum_{p,q=0}^{\infty} \left(-1\right)^{p+q}\frac{1}{p!q!}
    \sum_{\substack{\bfx_1,\ldots,\bfx_p \\
    \bfy_1,\ldots,\bfy_q}}
    \bar{\alpha}(\bfx_1) \ldots
    \bar{\alpha}(\bfx_n) \, \alpha(\bfy_1) \ldots \alpha(\bfy_q) \nonumber \\
    &&    ~~~~~~~~~\times \frac{\left\langle P(\bfx_1) \ldots P(\bfx_p) \,
      P^\dagger(\bfy_1) \ldots P^\dagger(\bfy_q) \right\rangle_c}{N^{2-p-q}} \comma
\label{eq:w_polyakov_series}
\end{eqnarray}
where the (properly normalised) connected expectation values are defined as:
\begin{flalign}
  & \frac{\left\langle P(\bfx_1) \ldots P(\bfx_p) \,
    P^\dagger(\bfy_1) \ldots P^\dagger(\bfy_q) \right\rangle_c}{N^{2-p-q}} \equiv \nonumber \\
  & \qquad \equiv
  \left.(-1)^{p+q+1} \frac{\delta^{p+q}}{\delta \bar{\alpha}(\bfx_1) \cdots \delta \bar{\alpha}(\bfx_p)
    \delta \alpha(\bfy_1) \cdots \delta \alpha(\bfy_q)}
  W(\alpha,\bar\alpha)\right|_{\alpha=\bar\alpha=0} 
\period
\end{flalign}
Finally, the Legendre transform of the generator of connected Polyakov
correlators yields the generator of 1PI diagrams:
\begin{subequations}
\begin{flalign}
& z(\bfx) = \frac{\delta}{\bar\alpha(\bfx)}
W(\alpha,\bar{\alpha}) = \frac{ \left\langle P(\bfx) \, \exp
    \left\{ - (\bar{\alpha} P + \alpha P^\dagger) \right\}
  \right\rangle }
{ \left\langle \exp
    \left\{ - (\bar{\alpha} P + \alpha P^\dagger) \right\} \right\rangle } \comma \\
& \Gamma(z,\bar{z}) = W(\alpha,\bar{\alpha}) - \left( \bar{\alpha} z +
  \alpha \bar{z} \right) \comma
\end{flalign}
\end{subequations}
where we have omitted the dependence on the spatial coordinates in
order to simplify the notation.  The functions
$z(\bar\alpha,\alpha),\bar z(\bar\alpha,\alpha)$ yield respectively
the expectation values of the Polyakov loop and its complex conjugate
in the presence of the external sources $\alpha,\bar\alpha$. We shall
refer to the function $\Gamma(z,\bar{z})$ as the quantum action for
the Polyakov loop. In general, it is different from the function
$V(z)$; however it can be shown that the two functions coincide for
space--independent Polyakov loops in the limit of infinite spatial
volume\footnote{
The provided definition of the quantum action and the decomposition of
$W(\alpha,\bar{\alpha})$ in connected expectation values are correct as long as
the vacuum is non--degenerate. Otherwise the relationship between
$(\alpha,\bar{\alpha})$ and $(z,\bar{z})$ is not invertible and the definition
of the quantum action is more involved; while a sum over all the vacua is
required in the Taylor expansion of $W(\alpha,\bar{\alpha})$. The non-degeneracy
condition is satisfied in the strong-coupling phase, in which we perform
explicit computation.
}.

The center is a symmetry group of the quantized theory if and only if
the effective potential $\Gamma(z,\bar{z})$ is invariant\footnote{ In
principle, one should consider the effective potential of a generic
loop arbitrarily wrapping along the time direction. This can be done
using the same formalism developed here. However we shall assume that
the Polyakov loop completely characterises the center symmetry.
}. Because of the Legendre transform, the effective potential is not
easily computed. However, since $z(u\alpha,\bar{u}\bar{\alpha}) = u
z(\alpha,\bar{\alpha})$, we can readily prove that, for any element of
the center $u$:
\begin{equation}
\Gamma(uz,\bar{u}\bar{z}) = W(u\alpha,\bar{u}\bar{\alpha}) -
  \bar{\alpha} z(\alpha,\bar{\alpha}) + \alpha
  \bar{z}(\alpha,\bar{\alpha})\comma
\end{equation}
and the following implications hold:
\begin{equation}
\textrm{center is a symmetry} \Leftrightarrow
\Gamma(uz,\bar{u}\bar{z})=\Gamma(z,\bar{z}) \Leftrightarrow
W(u\alpha,\bar{u}\bar{\alpha})=W(\alpha,\bar{\alpha}) 
\period
\end{equation}
Therefore, the functional $W(\alpha,\bar{\alpha})$ encodes all the
information that we need about the center symmetry. The terms of the
expansion in Eq.~(\ref{eq:w_polyakov_series}) are classified in
Tab.~\ref{tab:symexp} according to their symmetry properties.
\begin{table}[ht]
  \centering
  \begin{tabular}[h]{cc}
    Term in $W$ & Symmetry \\
    \hline
    $p = q \mod 2$ & $\Z_2$ \\
    $p = q \mod N$ & $\Z_{N}$ \\
    $p = q$ & $\mathrm{U}(1)$
  \end{tabular}
  \caption{Symmetry properties of the Polyakov connected correlators
  $\langle (P)^p (P^\dagger)^q \rangle_c$. We say that $G$ is a
  symmetry group for the Polyakov connected correlator if the latter
  is invariant under the transformation $\Omega(\bfx) \rightarrow u
  \Omega(\bfx)$, with $u \in G$. Note that $G$ is not necessarily a
  subgroup of the center.}
  \label{tab:symexp}
\end{table}

The connected correlators can be computed analytically for the pure
gauge theory in the strong coupling regime, where a topological
classification of diagrams can be setup which closely follows the one
obtained in the continuum by
't~Hooft~\cite{tHooft:1973jz,tHooft:2002yn}. The details relevant for
our computations are summarised in App.~\ref{app:a}. The large--$N$
behaviours of the correlators are:
\begin{flalign}
& \frac{\langle ( P )^p ( P^\dagger
    )^q \rangle_{c,\mathrm{YM}}}{N^{2-p-q}}= O(1)\comma \qquad \textrm{if } p = q \comma \nonumber\\
& \frac{\langle ( P )^p ( P^\dagger )^q \rangle_{c,\mathrm{YM}}}{N^{2-p-q}}
= O \left( \frac{1}{N}
\right)\comma \qquad \textrm{if } p = q \mod N \comma
\quad p \neq q \comma \nonumber \\
& \frac{\langle ( P )^p ( P^\dagger )^q
  \rangle_{c,\mathrm{YM}}}{N^{2-p-q}} = 0\comma \qquad 
\textrm{if } p \neq q \mod N \period
\end{flalign}

\section{The connected graphs in the large--mass expansion}
\label{sec:connected}

Let us now discuss the introduction of fermions in the theory. Since
we are interested in the orientifold planar equiavalence, we shall
consider fermions in arbitrary representations. The fermionic
effective action can be written as a sum of Wilson loops in the
large--mass phase~\cite{Wilson:1974sk}:
\begin{equation}
S_f(U_{k}, \Omega) = -\sum_{\omega \in \mathcal{C}}
c(\omega) \, \tr \R[U(\omega) ] 
\comma
\end{equation}
where $\omega$ indicates a generic closed path on the lattice of
length $L(\omega)$, $U(\omega)$ is the parallel transport along
$\omega$, and $\R[U]$ is the matrix representing $U$ in the
representation $\R$. The coefficients $c(\omega)$ depend on the spin
structure of the chosen discretisation of the Dirac operator and are
independent of the colour structure. They are of order of
$m^{-L(\omega)}$, where $m$ is the bare fermion mass. It will be
convenient to split the set $\mathcal{C}$ of all the closed paths into
the union of the sets $\mathcal{C}(w)$ of closed paths with winding
number $w$ around the thermal dimension.

The connected expectation value of Polyakov loops can be formally
thought as a function of the coefficients $c(\omega)$. In the
large--mass phase, it can be expanded as a power series around
$c(\omega)=0$:
\begin{equation}
\frac{1}{ N^{2-p-q} } \langle (P)^p ( P^\dagger )^q \rangle_c =
\sum_{n=0}^{\infty} \frac{1}{n!} \sum_{\omega_1 \dots \omega_n}
\frac{c(\omega_1) \cdots c(\omega_n)}{ N^{2-p-q} } \left.
\frac{ d^n \langle (P)^p ( P^\dagger )^q \rangle_c }
{dc(\omega_1) \cdots dc(\omega_n)} \right|_{S_f=0} \period
\label{eq:P_large_mass}
\end{equation}
Each derivative with respect to $c(\omega)$ corresponds to the
insertion of a vertex $ \tr \R[U(\omega)]$ in the expectation
value. Since the derivative must be computed at $S_f=0$, each term in
the expansion above can be written as combinations of connected
expectation values in the pure Yang--Mills, as shown in some explicit
examples later on.  Therefore the asymptotic behaviour of the
connected expectation values in the theory with dynamical fermions can
be reconstructed from the asymptotic behaviour of the connected
expectation values in the pure Yang--Mills.

Let $\omega_1, \dots \omega_n$ be some closed paths and $w_1, \dots
w_n$ be their respective winding numbers around the thermal direction.
Consider the generic normalised connected expectation value in the
pure Yang--Mills theory:
\begin{equation} \label{eq:YM_general}
\frac{
\langle \prod_i \tr U(\omega_i) \rangle_{c,\mathrm{YM}}
}{N^{2-n}} \period
\end{equation}
This quantity is always finite in the large--$N$ limit (it could be
also $0$ and in this case we say that the leading order is not
saturated). This fact is proved in the strong--coupling, and in the
perturbative expansion, and it is consistent with results from
numerical simulations.  Moreover, it is equivalent to the $N^2$
scaling of the free energy of the system, when external sources of
order $N$ are coupled to the operators $\tr U(\omega_i)$ in analogy to
the definition~\eqref{eq:Walpha}, and consequently it is equivalent to
the $N^2$ scaling of the quantum action for the expectation values of
the operators $\tr U(\omega_i)/N$. The following cases can arise:
\begin{itemize}
\item $\sum_i w_i \neq 0 \mod N$. The expression in
Eq.~(\ref{eq:YM_general}) transforms non--trivially under center
transformations. Since the center symmetry is not broken in the
confined phase, the expression is exactly zero for every value of $N$.
\item $\sum_i w_i = 0 \mod N$ but $\sum_i w_i \neq 0$ (see
Figs.~\ref{fig:ym2}, \ref{fig:ym3}). The expression in
Eq.~(\ref{eq:YM_general}) is invariant under $Z_N$ but not
$\mathrm{U}(1)$. Using the strong--coupling expansion (see
App.~\ref{app:a}) it is possible to prove that the correlator vanishes
in the large--$N$ limit. This result can be extended beyond the
strong--coupling phase, under the commonly accepted assumption that
the $\mathrm{U}(N)$ and $\mathrm{SU}(N)$ Yang--Mills theories differ
in the large--$N$ limit only for subleading contributions. If the
gauge group were $\mathrm{U}(N)$, the expression in
Eq.~(\ref{eq:YM_general}) would be exactly zero for every value of
$N$. Since we are interested in the $\mathrm{SU}(N)$ gauge group, it
must be zero in the large--$N$ limit.
\item The loops with winding number different from zero appear in
pairs with opposite winding number, as shown in Fig.~\ref{fig:ym1}.
In particular $\sum_i w_i = 0$. From the strong--coupling expansion
(see App.~\ref{app:a}) we get that the expression in
Eq.~(\ref{eq:YM_general}) in this case is different from zero in the
large--$N$ limit. We will assume that this result extends beyond the
strong--coupling phase.
\item $\sum_i w_i = 0$ but some of the loops with winding number $w
\neq 0$ cannot be paired with a loop with winding number $-w$.
Although the expression in Eq.~(\ref{eq:YM_general}) is invariant
under the center (both $\Z_N$ and $\mathrm{U}(1)$), it vanishes in the
strong--coupling phase (see App.~\ref{app:a}) because of the topology
of the torus. We cannot say if this result extends beyond the
strong--coupling expansion, but anyway we will not use it in this
work.
\end{itemize}
%

%\begin{figure}[!ht]
\FIGURE{
  \centering
  \epsfig{file=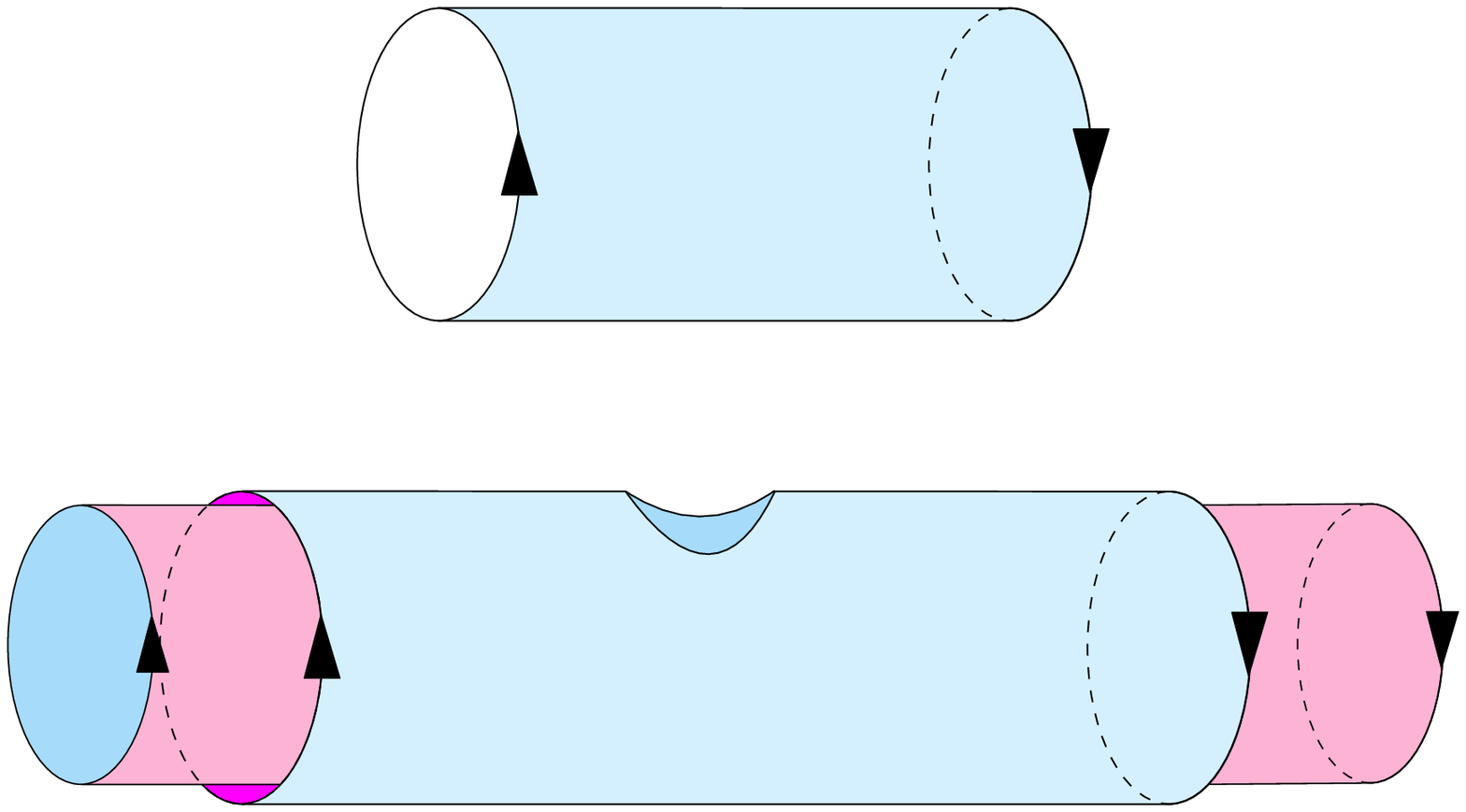,scale=0.8}  
  \caption{Representations of some connected expectation values of
    products of loops with net winding number equal to zero, in pure
    Yang--Mills. The first one is $\langle \tr \Omega \ \tr
    \Omega^\dagger \rangle_{c,\mathrm{YM}} $; the Polyakov loops are
    connected by an oriented surface that wraps around the thermal
    dimension. In the strong--coupling expansion, the surface is tiled
    by plaquettes coming from the Wilson action; products of two (or
    more) group elements associated to each link of the lattice are
    integrated with respect to the Haar measure. This graph yields a
    contribution $O(N^0)$. The second graph represents $\langle (\tr
    \Omega \ \tr \Omega^\dagger)^2 \rangle_{c,\mathrm{YM}} $; the two
    tubes are glued together through a hole. This graph is
    $O(N^{-2})$.}
  \label{fig:ym1}
}
%\end{figure}

%\begin{figure}[!ht]
\FIGURE{
  \centering
  \epsfig{file=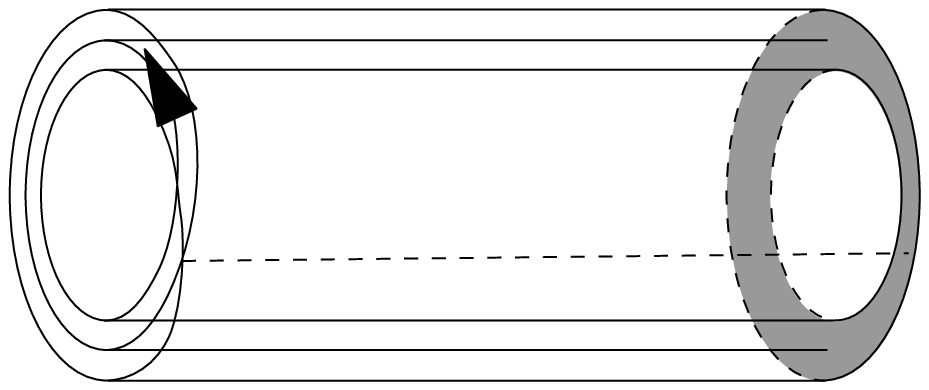,scale=1}  
  \caption{Graphical representation of $\langle \tr (\Omega^N)
    \rangle_{c,\mathrm{YM}} $ for $N=3$. The surface wraps $N$ times
    around the thermal direction and ends in the gray area, which
    represents the integration of the product of $N$ group elements,
    all oriented in the same direction. This integration yields a
    non--zero value, because the product of $N$ fundamental
    representations of $\mathrm{SU}(N)$ contains a singlet, given by the
    contraction with the completely skew--symmetric tensor. This graph
    yields a contribution $O(N^0)$.}
  \label{fig:ym2}
}
%\end{figure}

%\begin{figure}[!ht]
\FIGURE{
  \centering
  \epsfig{file=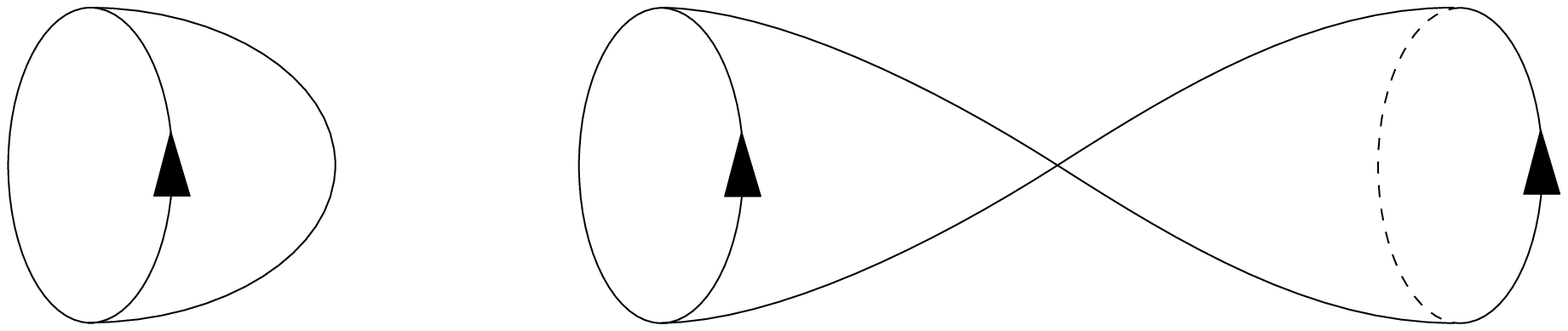,scale=0.8}  
  \caption{These graphs are the would--be leading contributions of
    respectively $\langle \tr \Omega \rangle_{\mathrm{YM}} $ and
    $\langle (\tr \Omega)^2 \rangle_{c,\mathrm{YM}} $. They are
    forbidden by the topology of the thermal dimension.}
  \label{fig:ym3}
}
%\end{figure}

It is worth to remind that in the theories with fermions the expansion
of the fermionic effective action in Eq.~\eqref{eq:P_large_mass} is
convergent only in the limit where the bare fermion mass is
large, and therefore yields only limited information on the continuum
limit of the lattice theory. It provides nonetheless a framework where
analytical calculations can be performed and the planar equivalence
can be tested explicitly. Note that for fermions in the adjoint
representation, the full theory is still invariant under the center
group $\Z_N$ and the large--$N$ behaviour of the connected
correlators is the same as the one we obtained above for the pure
gauge theory.

\subsection{Fundamental fermions}
\label{subsec:connected:fundamental}

When fermions in the fundamental representation are considered, the
generic term of the r.h.s. of the Eq.~(\ref{eq:P_large_mass}) is:
\begin{equation}
\label{eq:fundf}
\frac{1}{ N^{2-p-q} } \left. \frac{ d^n \langle (P)^p ( P^\dagger )^q \rangle_{c,\mathrm{F}} }
{dc(\omega_1) \cdots dc(\omega_n)} \right|_{S_f=0} =
\frac{
\langle ( P )^p ( P^\dagger )^q\, \tr U(\omega_1) \ldots  
\tr U(\omega_n) \rangle_{c,\mathrm{YM}}
}{ N^{2-p-q} } \period
\end{equation}
The expectation value that appears in the numerator of the r.h.s. of
Eq.~(\ref{eq:fundf}) is at most of order $N^{2-p-q-n}$, since it
contains $p+q+n$ loops. Thus for $n \neq 0$, the whole term is at most
of order $N^{-1}$. Now, if $p=q$ the Yang--Mills contribution ($n=0$)
is $O(1)$ and dominates the sum. Otherwise, the leading contribution
is at most of order $N^{-1}$:
\begin{flalign}
& \frac{ \langle ( P )^p ( P^\dagger )^q \rangle_{c,\mathrm{F}} }{
  N^{2-p-q} } \simeq \frac{ \langle ( P )^p ( P^\dagger )^q \rangle_{c,\mathrm{YM}} }{
  N^{2-p-q} } = O(1)\comma \qquad \textrm{if } p = q \nonumber \\
& \frac{ \langle ( P )^p ( P^\dagger )^q \rangle_{c,\mathrm{F}} }{
  N^{2-p-q} } = O \left( \frac{1}{N} \right)\comma
\qquad 
\textrm{if } p \neq q \period
\end{flalign}

For $p \neq q$, the $1/N$ scaling is only an upper limit for the
asymptotic behaviour. However in the case of the expectation value of
the Polyakov loop, the $O(N^{-1})$ is saturated by the contribution:
\begin{flalign}
\frac{ \langle P \rangle }{N} = & \sum_{n=0}^{\infty} \frac{1}{n!}
\sum_{\omega_1 \dots \omega_n} \frac{c(\omega_1) \cdots c(\omega_n)}{
N^{2-p-q} } \left.  \frac{ d^n \langle P \rangle_c } {dc(\omega_1)
\cdots dc(\omega_n)} \right|_{S_f=0} = \nonumber \\ = & \frac{1}{N}
\sum_{\omega \in \mathcal{C}(-1)} c(\omega) \langle P \ \tr U(\omega)
\rangle_{c,\mathrm{YM}} + O \left( \frac{1}{N^2} \right) \period
\end{flalign}
We recall that $\mathcal{C}(-1)$ is the set of the closed paths on the
lattice with winding number equal to $-1$ around the thermal
dimension.

From the large--$N$ behaviour of the coefficients of the Taylor
expansion, we see that, in the large--$N$ limit, the functional $W$ is
the same for Yang--Mills and for the theory with fundamental quarks:
\begin{flalign}
& \lim_{N \rightarrow \infty} W_{F}(\alpha,\bar{\alpha}) = \lim_{N
  \rightarrow \infty} W_{YM}(\alpha,\bar{\alpha}) 
= \nonumber \\
& \qquad = 1 - \sum_{p=0}^{\infty} \frac{|\alpha|^{2p}}{(p! 2^p)^2}
\lim_{N \rightarrow \infty} \frac{ \langle \left| \tr 
\Omega \right|^{2p} \rangle_c }{ N^{2-2p} } \period
\end{flalign}
This is the manifestation in this particular sector of the theory of
the usual subleading contribution from the fermion determinant when
fermions are in the fundamental representation.

\subsection{S/AS fermions}
\label{subsec:connected:sas}

Again, in the generic term of the r.h.s. of the
Eq.~(\ref{eq:P_large_mass}):
\begin{equation}
\frac{1}{ N^{2-p-q} } \left. \frac{ d^n \langle (P)^p ( P^\dagger )^q \rangle_{c,\mathrm{S/AS}} }
{dc(\omega_1) \cdots dc(\omega_n)} \right|_{S_f=0} \comma
\end{equation}
each derivative in $c(\omega)$ produces the insertion of a vertex $\tr
\R[U(\omega)]$. When the fermions are in the (anti)symmetric
representation, we can use the algebraic relationships:
\begin{equation}
\tr \mathrm{S/AS}[U] = \frac{(\tr U)^2 \pm \tr (U^2)}{2} \period
\end{equation}

In general, the insertion of some term of the form $(\tr U)^2 /2$ will
disconnect the expectation value $\langle (P)^p ( P^\dagger )^q
\rangle_{c,\mathrm{S/AS}}$. In order to gain some confidence with
these computations, we report in the following subsections the
explicit computation of the leading orders of $ \langle (P)^2
\rangle_{c,\mathrm{S/AS}} $, $ \langle P P^\dagger
\rangle_{c,\mathrm{S/AS}} $ and $ \langle P \rangle_{c,\mathrm{S/AS}}
$. The end of the section is devoted to a discussion of the general
case.

Before entering into the details of the computations, we summarise
here the results:
\begin{flalign}
& \frac{ \langle ( P )^p ( P^\dagger )^q \rangle_{c,\mathrm{S/AS}} }{
  N^{2-p-q} } = O(1) \comma
\qquad \textrm{if } p - q \textrm{ even} \comma \nonumber \\
& \frac{ \langle ( P )^p ( P^\dagger )^q \rangle_{c,\mathrm{S/AS}} }{
  N^{2-p-q} } = O\left( \frac{1}{N} \right) \comma \qquad \textrm{if } p - q \textrm{ odd} 
\textrm{ and } N\textrm{ odd} \comma \nonumber \\
& \frac{ \langle ( P )^p ( P^\dagger )^q \rangle_{c,\mathrm{S/AS}} }{
  N^{2-p-q} } = 0 \comma\qquad 
\textrm{if } p - q \textrm{ odd} \textrm{ and } N \textrm{ even} \period
\end{flalign}
Hence the large--$N$ limit for the functional $W$ yields:
\begin{flalign}
& \lim_{N \rightarrow \infty} W_{S/AS}(\alpha,\bar{\alpha}) = \nonumber \\
& \qquad = 1 - \sum_{\substack{p,q=0\\p-q \textrm{ even}}}^{\infty}
\frac{1}{p!q!} \bar{\alpha}^p \alpha^q
\lim_{N \rightarrow \infty} \frac{ \langle ( P )^p ( P^\dagger )^q
  \rangle_{c,\mathrm{S/AS}} }
{ N^{2-p-q} } \period
\end{flalign}
Since all the terms with even $(p-q)$ contribute to the sum, we get
$$W_{S/AS}(\alpha,\bar{\alpha}) = W_{S/AS}(u \alpha,\bar{u}
\bar{\alpha})$$
in the planar limit if and only if $u=\pm 1$. Therefore
a $\Z_2$ symmetry is recovered in the large--$N$ limit.

\subsubsection{Computation of $ \langle P \rangle_{\mathrm{S/AS}} $}

Let us start with the single--derivative term in the
expansion~(\ref{eq:P_large_mass}):
\begin{flalign}
  \frac{1}{N} \left. \frac{ d \langle P \rangle_{\mathrm{S/AS}}
    }{dc(\omega)} \right|_{S_f=0} = & \frac{1}{N} \left\{ \langle P\
    \tr \R[U(\omega)] \rangle_{\mathrm{YM}} - \langle P
    \rangle_{\mathrm{YM}} \langle \tr \R[U(\omega)]
    \rangle_{\mathrm{YM}} \right\} = \nonumber \\ = & \frac{1}{2N}
    \left\{ \langle P\ [\tr U(\omega)]^2 \rangle_{\mathrm{YM}} \pm
    \langle P\ \tr [U(\omega)^2] \rangle_{\mathrm{YM}} \right\} =
    \nonumber \\ = & \frac{1}{2N} \left\{ \langle P\ [\tr U(\omega)]^2
    \rangle_{c,\mathrm{YM}} + 2 \langle P\ \tr U(\omega)
    \rangle_{c,\mathrm{YM}} \langle \tr U(\omega)
    \rangle_{c,\mathrm{YM}} \right.\nonumber \\ & ~~~~~~~~\left.\pm
    \langle P\ \tr [U(\omega)^2] \rangle_{c,\mathrm{YM}} \right\}
    \period
\end{flalign}
Let $w$ be the winding number around the thermal dimension of the
closed path $\omega$. We shall analyse each term in turn. The term
$\langle P\ [\tr U(\omega)]^2 \rangle_{c,\mathrm{YM}}$ saturates its
$N^{-1}$ behaviour only if $1+2w=0$, corresponding to correlators that
are invariant under the action of $\mathrm{U}(1)$. But this equation has no
integer solution, therefore the highest order is never saturated. The
only non--vanishing contributions come from loops that satisfy
$1+2w=kN$ (both $k$ and $N$ must be odd). In this case, the first term
goes like $N^{-2}$.

The second term $\langle P\ \tr U(\omega) \rangle_{c,\mathrm{YM}}
\langle \tr U(\omega) \rangle_{c,\mathrm{YM}}$ saturates its $N$
asymptotic behaviour only if each factor is separately invariant under
the action of $\mathrm{U}(1)$. This corresponds to the equations $1+w=0$ and
$w=0$. Again, this equation admits no solution and the leading order
is not saturated. We can look for subleading contributions. Since we
want $\langle \tr U(\omega) \rangle_{c,\mathrm{YM}}$ not vanishing, it
must be $w=kN$. Requiring that $\langle P\ \tr U(\omega)
\rangle_{c,\mathrm{YM}}$ is not vanishing, we get $1+w=1+kN=k'N$ which
implies $(k'-k)N=1$. This equation has no solution, and therefore the
second term is always zero.

The condition for the last term $\langle P\ \tr [U(\omega)^2]
\rangle_{c,\mathrm{YM}}$ to saturate its $N^0$ behaviour is again
$1+2w=0$. Looking for subleading contributions, we have to request
$1+2w=kN$. This equation admits a solution if both $k$ and $N$ are
odd, as for the first term. But unlike the first term, the last one
goes like $N^{-1}$.

Putting all these contributions together, the single--derivative in
Eq.~(\ref{eq:P_large_mass}) is of order $N^{-2}$ and this behaviour is
summarised by the formula:
\begin{flalign}
  \frac{1}{N} \left. \frac{ d \langle P \rangle_{\mathrm{S/AS}}
    }{dc(\omega)} \right|_{S_f=0} = & \pm \frac{1}{2N} \langle P\ \tr
  [U(\omega)^2] \rangle_{c,\mathrm{YM}} + O \left( \frac{1}{N^3}
  \right) \period
\end{flalign}

Consider now a generic term in the
expansion~(\ref{eq:P_large_mass}). We will not write the explicit
expression, but it should be clear that it can be written as a sum of
products of connected expectation values in the pure Yang--Mills
theory. Let $w_i$ be the winding number of the path $\omega_i$.  Only
products that are invariant under the action of $\mathrm{U}(1)$ saturate the
leading contribution; hence the sum of all the winding numbers must
vanish:
\begin{equation}
1+2 \sum_i w_i = 0 \period
\end{equation}
The factor $2$ is due to the two--index representation. For each closed
path $\omega_i$ with winding number $w_i$, a loop $\tr \R[U(\omega)]$
with $N$--ality equal to $2w_i$ is inserted. The equation above has no
solution, therefore the leading order for $\langle P
\rangle_{c,\mathrm{S/AS}}$ is not saturated. This argument lets us to
conclude that:
\begin{equation}
\lim_{N\rightarrow\infty} \frac{\langle P \rangle_{\mathrm{S/AS}}}{N}
= 0 \comma
\end{equation}
but we still cannot say if it is a $O(1/N)$ or rather a $O(1/N^2)$ as
the single--derivative term suggests. Since this is not essential to
our discussion of the center symmetry, we state simply the result as:
\begin{flalign}
\frac{\langle P \rangle}{N} = &
\pm \frac{1}{N} \sum_{n=0}^{\infty} \frac{1}{n!} \sum_{k \textrm{ odd}}
\sum_{\bar{\omega} \in \mathcal{C}\left( \frac{kN-1}{2} \right)}
\sum_{\omega_1 \dots \omega_n} \nonumber \\
& c(\bar{\omega}) c(\omega_1) \cdots c(\omega_n) \left.
\frac{ d^n \langle P \tr [ U(\bar{\omega})^2 ] \rangle_c }
{dc(\omega_1) \cdots dc(\omega_n)} \right|_{S_f=0} + O\left( \frac{1}{N^3} \right) \comma
\end{flalign}
which means that $\langle P \rangle/N$ is of order $O(1/N^2)$.

\subsubsection{Computation of $ \langle (P)^2 \rangle_{c,\mathrm{S/AS}} $}

We want to check here that the leading term of $ \langle (P)^2
\rangle_{c,\mathrm{S/AS}} $, which is expected to be $N^0$, is
actually different from zero. This can be easily shown by computing
the single--derivative term in the expansion
Eq.~(\ref{eq:P_large_mass}). Starting from the definition of the
connected correlator: $ \langle P^2 \rangle_{c,\mathrm{S/AS}} =
\langle P^2 \rangle_{\mathrm{S/AS}} - \langle P
\rangle_{\mathrm{S/AS}}^2 $, and assuming that $\omega$ is a generic
closed path with winding number $-1$ around the thermal dimension, we
can rewrite the single--derivative coefficient of the expansion of
$\langle P^2 \rangle_{c,\mathrm{S/AS}}$ as:
\begin{flalign}
  & \left. \frac{ d \langle P^2 \rangle_{c,\mathrm{S/AS}}
    }{dc(\omega)} \right|_{S_f=0} = \langle P^2 \R[U(\omega)]
    \rangle_{\mathrm{YM}} - \langle P^2 \rangle_{\mathrm{YM}} \langle
    \R[U(\omega)] \rangle_{\mathrm{YM}} = \nonumber \\ & \qquad =
    \frac{1}{2} \langle P^2 [\tr U(\omega)]^2 \rangle_{\mathrm{YM}} =
    \nonumber \\ & \qquad = \frac{1}{2} \left\{ \langle P^2 [\tr
    U(\omega)]^2 \rangle_{c,\mathrm{YM}} + 2 \langle P \ \tr U(\omega)
    \rangle_{c,\mathrm{YM}} \langle P \ \tr U(\omega)
    \rangle_{c,\mathrm{YM}} \right\} = \nonumber \\ & \qquad = \langle
    P \ \tr U(\omega) \rangle_{c,\mathrm{YM}} \langle P \ \tr
    U(\omega) \rangle_{c,\mathrm{YM}} + O\left( \frac{1}{N^2} \right)
    \comma
\end{flalign}
as illustrated in Fig.~\ref{fig:sas_p2}. This is exactly a $O(N^0)$ term.
%
%\begin{figure}[!ht]
\FIGURE{
  \centering
  \epsfig{file=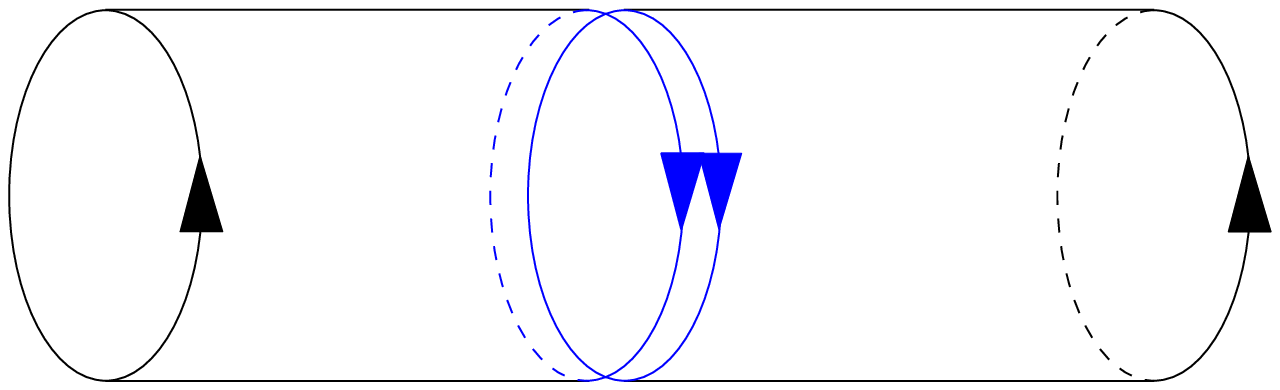,scale=0.8}  
  \caption{This graph represents a leading contribution of $\langle
    (\tr \Omega)^2 \rangle_{c,\mathrm{S/AS}} $. The pair of blue loops
    represent the insertion of a vertex $ \tr \R[U(\omega)]$, due to
    the derivative with respect to $c(\omega)$.}
  \label{fig:sas_p2}
}
%\end{figure}

\subsubsection{Computation of $ \langle P P^\dagger \rangle_{c,\mathrm{S/AS}} $}
This is the easiest example: since $P P^\dagger$ is invariant under
the action of $\mathrm{U}(1)$, the zero--derivative term $ \langle P
P^\dagger \rangle_{c,\mathrm{YM}} $ in the
expansion~(\ref{eq:P_large_mass}) trivially saturates the $N^0$
behaviour of $ \langle P P^\dagger \rangle_{c,\mathrm{S/AS}} $.

\subsubsection{The generic $ \langle P^p (P^\dagger)^q \rangle_{c,\mathrm{S/AS}} $}

As already discussed in the case of $ \langle P
\rangle_{c,\mathrm{S/AS}} $, each term in the
expansion~(\ref{eq:P_large_mass}) of the generic connected expectation
value can be written as a sum of products of connected expectation
values with respect the YM vacuum. Each of these connected expectation
values contains some $P$'s and $P^\dagger$'s from the original
expectation value, and some $\tr U(\omega)$ from the derivatives with
respect to the coefficient $c(\omega)$. When the derivative with
respect to $c(\omega_i)$ is computed, a $[\tr U(\omega_i)]^2$ or a
$\tr [U(\omega_i)^2]$ is inserted. In both cases, if $w_i$ is the
winding number of the path $\omega_i$, a parallel transport with
winding number $2w_i$ is inserted. A necessary condition for the
leading order to be saturated is that the overall winding number, that
is the sum of all the winding numbers of the operators involved, must
be zero (this corresponds to the invariance under action of
$\mathrm{U}(1)$):
\begin{equation}
p-q+2 \sum_i w_i = 0 \period
\end{equation}
This argument implies that $p-q$ must be even. If it is odd, the
leading behaviour cannot be saturated.

Now we want to show that this is also a sufficient condition, by
explicitly constructing a term that saturates the leading
behaviour. The $p=q$ case is trivial since the leading behaviour is
saturated by the pure YM term:
\begin{equation}
\frac{ \langle ( P )^p ( P^\dagger )^p \rangle_{c,\mathrm{S/AS}} }{ N^{2-2p} } =\frac{ \langle ( P )^p ( P^\dagger )^p \rangle_{c,\mathrm{YM}} }{ N^{2-2p} } + \dots
\end{equation}
Consider now the $p>q$ case (the opposite can be obtained by
charge--conjugation). If $\omega$ is a path with winding number $-1$, a
leading contribution comes for instance from (see fig.~\ref{fig:sas_generic}):
\begin{flalign}
  \frac{1}{ N^{2-p-q} }
  & \left. \frac{ d^n \langle (P)^p ( P^\dagger )^q \rangle_c }{dc(\omega)^n} \right|_{S_f=0} = \nonumber \\
  & = \frac{ p! (p-q)! }{ \left( \frac{p+q}{2} \right) ! \left[ \left(
        \frac{p-q}{2} \right) ! \right]^2 } \frac{ \langle ( P
    )^{\frac{p+q}{2}} [\tr U(\omega)]^{\frac{p-q}{2}} ( P^\dagger )^q
    \rangle_{c,\mathrm{YM}} }{ N^{2-p-q} } ( \langle P \ \tr U(\omega)
  \rangle_{c,\mathrm{YM}} )^{\frac{p-q}{2}} + \dots
\end{flalign}

%\begin{figure}[!ht]
\FIGURE{
  \centering
  \epsfig{file=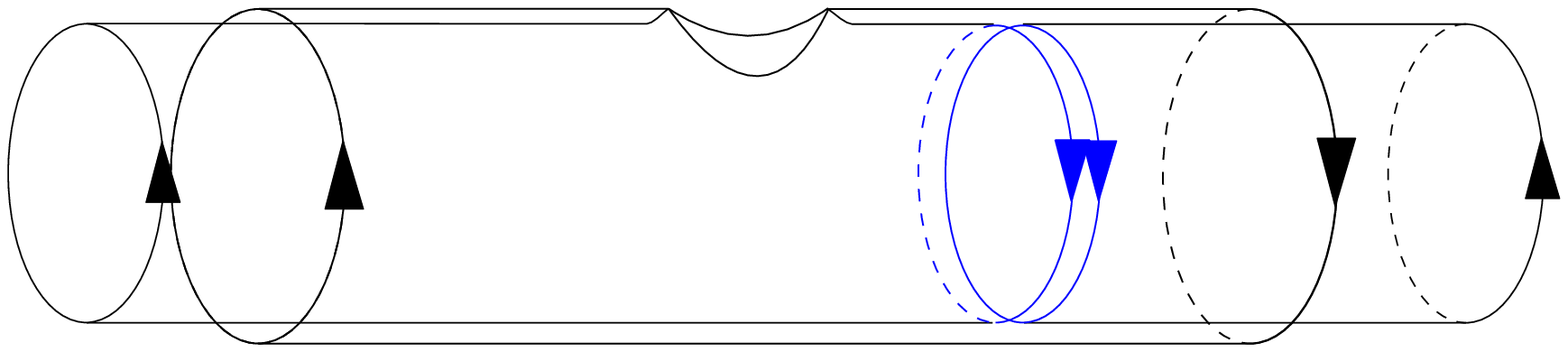,scale=0.8}  
  \caption{This graph is a leading contribution of $\langle (\tr
    \Omega)^3 \tr \Omega^\dagger \rangle_{c,\mathrm{S/AS}} $. The pair
    of blue loops represent the insertion of a vertex $ \tr
    \R[U(\omega)]$, due to the derivative with respect to
    $c(\omega)$.}
  \label{fig:sas_generic}
}
%\end{figure}

\section{The effective potentials in the large--mass expansion}
\label{sec:effpotential}

The analysis of the center symmetry of the effective potential can be
deduced entirely from the results presented in the previous
section. Nonetheless, it is interesting to compare those results with
the prediction in Ref.~\cite{Sannino:2005sk} about the effective
potential of the Polyakov loop, bearing in mind that in
Ref.~\cite{Sannino:2005sk} only the lowest--dimensional terms
(relevant operators) are kept in the effective potential for the
Polyakov loop $\Gamma(z,\bar z)$. They can be summarised as follows.

\begin{description}

\item[Pure gauge or adjoint fermions.]
\begin{equation}
\Gamma(z,\bar{z}) = a_2 z\bar{z} + a_4 ( z\bar{z} )^2 + a_N ( z^N + \bar{z}^N )
\end{equation}

\item[Fundamental fermions.]
\begin{equation}
\Gamma(z,\bar{z}) = b_1 ( z + \bar{z} ) + a_2 z\bar{z} + a_4 (
z\bar{z} )^2 + 
a_N ( z^N + \bar{z}^N )
\end{equation}

\item[S/AS fermions with $N$ even.]
\begin{equation}
\Gamma(z,\bar{z}) = b_2 ( z^2 + \bar{z}^2 ) + a_2 z\bar{z} + a_4 (
z\bar{z} )^2 + 
a_N ( z^N + \bar{z}^N )
\end{equation}

\item[S/AS fermions with $N$ odd.]
\begin{equation}
\Gamma(z,\bar{z}) = b_1 ( z + \bar{z} ) + b_2 ( z^2 + \bar{z}^2 ) +
a_2 z\bar{z} + 
a_4 ( z\bar{z} )^2 + a_N ( z^N + \bar{z}^N )
\end{equation}

\end{description}

In the large--$N$ limit not all the coefficients survive. In particular:
\begin{equation}
\lim_{N \rightarrow \infty} a_N(N) = 0 \comma
\end{equation}
for all the theories; moreover, in the case of fundamental fermions:
\begin{equation}
\lim_{N \rightarrow \infty} b_1(N) = 0 \comma
\end{equation}
while in the case of S/AS fermions:
\begin{equation}
\lim_{N \rightarrow \infty} b_1(N) = 0 \comma \qquad \qquad \lim_{N \rightarrow \infty} b_2(N) \neq 0 \comma
\end{equation}

As shown in the previous Section, this behaviour is constrained by the
symmetry properties of each term in the potential. It is nonetheless
intersting to check explicitly the last two equations.

\subsection{The $b_1$ coefficient}
\label{subsec:effpotential:b1}

Since $\alpha(z,\bar{z}) = -2 \bar{\partial}_z \Gamma(z,\bar{z})$,
from the Taylor expansion of the effective potential:
\begin{equation}
b_1 = \partial_z \Gamma(0,0) = - \frac{1}{2} \alpha_0 \comma
\end{equation}
where $\alpha_0$ is the real source, defined by the condition
$z(\alpha_0,\alpha_0) = \bar{z}(\alpha_0,\alpha_0) = 0$, which is the
same as:
\begin{equation}
(\partial_\alpha + \bar{\partial}_\alpha ) W(\alpha_0,\alpha_0) = 0 \period
\end{equation}
Using the series expansion of $W(\alpha, \bar{\alpha})$, the above
condition becomes:
\begin{equation}
\frac{\langle \tr \Omega \rangle}{N} - \alpha_0
\sum_{n=0}^{\infty} 
\frac{(- \alpha_0)^n}{(n+1)!} \frac{\langle \left( \Real 
\tr \Omega \right)^{n+2} \rangle_c}{N^{-n}} = 0 \period
\end{equation}
Since $\frac{\langle \tr \Omega \rangle}{N}$ vanishes in the
large--$N$ limit, the equation can be solved iteratively. At the
leading order:
\begin{equation}
\alpha_0 = \frac{ \langle \tr \Omega \rangle }{ N \langle
  \left( \Real \,
\tr \Omega \right)^2 \rangle_c } + \textrm{ subleadings} \period
\end{equation}
Expanding the denominator, the $b_1$ coefficient is:
\begin{equation}
b_1 = - \frac{ \langle \tr \Omega \rangle }{ N \left( \langle
    \left| \tr \Omega \right|^2 \rangle_c + \langle \left(
      \tr \Omega \right)^2 \rangle_c \right) } 
+ \textrm{ subleadings} \comma
\end{equation}
and it vanishes in the large--$N$ limit, with both fundamental and
S/AS fermions.

\subsection{The $b_2$ coefficient}
\label{subsec:effpotential:b2}

The second order in the Taylor expansion of the effective potential is
obtained by inverting the Hessian matrix of the functional $W$:
\begin{equation}
\left( \begin{array}{cc}
2 b_2 & a_2 \\
a_2 & 2 b_2
\end{array} \right) (0) = 
- \frac{1}{4} \left( \begin{array}{cc}
\bar{\partial}^2_\alpha W & \partial_\alpha \bar{\partial}_\alpha W \\
\partial_\alpha \bar{\partial}_\alpha W & \partial^2_\alpha W
\end{array} \right)^{-1} (\alpha_0) \period
\end{equation}
The entries of the Hessian matrix in the large--$N$ limit are:
\begin{flalign}
\partial^2_\alpha W(\alpha_0) = & -\frac{1}{4}
\frac{ \langle (\tr \Omega)^2 \; \exp \left\{ - N \alpha_0
    \Real \tr \Omega \right\} \rangle }{ \langle \exp
  \left\{ - N \alpha_0 \Real \tr 
\Omega \right\} \rangle } + \nonumber \\
& + \frac{1}{4} \left( \frac{ \langle \tr \Omega \; \exp
    \left\{ - N \alpha_0 \Real \tr \Omega \right\}
    \rangle }{ \langle \exp \left\{ - N \alpha_0 \Real
      \tr 
\Omega \right\} \rangle } \right)^2 = \nonumber \\
= & -\frac{1}{4} \langle (\tr \Omega)^2 \rangle_c + \dots
\end{flalign}
\begin{flalign}
\partial_\alpha \bar{\partial}_\alpha W(\alpha_0) = & - \frac{1}{4}
\frac{ \langle |\tr \Omega|^2 \; \exp \left\{ - N \alpha_0
    \Real \tr \Omega \right\} \rangle }{ \langle \exp
  \left\{ - N \alpha_0 \Real 
\tr \Omega \right\} \rangle } + \nonumber \\
& + \frac{1}{4} \left| \frac{ \langle \tr \Omega \; 
\exp \left\{ - N \alpha_0 \Real \tr \Omega \right\}
\rangle }
{ \langle \exp \left\{ - N \alpha_0 \Real \tr \Omega
  \right\} 
\rangle } \right|^2 = \nonumber \\
= & -\frac{1}{4} \langle |\tr \Omega|^2 \rangle_c + \dots
\end{flalign}
Computing the inverse of the Hessian yields:
\begin{flalign}
& a_2 = - \frac{ \langle |\tr \Omega|^2 \rangle_c }
{ \left[ \langle (\tr \Omega)^2 \rangle_c \right]^2 - 
\left[ \langle |\tr \Omega|^2 \rangle_c \right]^2 } \comma \\
& b_2 = \frac{ \langle (\tr \Omega)^2 \rangle_c }
{ 2 \left[ \langle (\tr \Omega)^2 \rangle_c \right]^2 - 
\left[ \langle |\tr \Omega|^2 \rangle_c \right]^2 } \period
\end{flalign}
In particular, the $a_2$ coefficient is always of order 1; the $b_2$
coefficient is of order $N^{-1}$ for fundamental fermions, while it is
of order 1 for S/AS fermions.

\section{The effective potential in the deconfined phase}
\label{sec:hitemp}

We consider here the high--temperature regime. In other words, the
space is taken to be $\mathbb{R}^3 \times S_1$, where the extension
$L$ of the compact dimension is much smaller than
$\Lambda_{\mathrm{QCD}}^{-1}$. Antiperiodic boundary conditions
conditions are imposed for the fermion fields, so that the path
integral can be interpreted as the partition function for a quantum
field theory at finite temperature. 

If $e^{iv_1}, \dots e^{iv_N}$ are the eigenvalues of the parallel
transport $\Omega$ around the compact dimension, the effective
potential for $v_1, \dots v_N$ can be computed in the one--loop
approximation. We refer to Ref.~\cite{Barbon:2005zj} for the details
of the computation. At fixed $N$, the effective potential for
respectively OrientiQCD and AdjQCD are:
\begin{flalign}
  & V_{\mathrm{Or}}(v) = \frac{1}{L^4} \left[ \sum_{i,j=1}^{N}
  f(0,v_i-v_j) - 2 N_f \sum_{i<j=1}^{N} f(mL,v_i+v_j+\pi) \right]
  \comma \\ & V_{\mathrm{Adj}}(v) = \frac{1}{L^4} \left[
  \sum_{i,j=1}^{N} f(0,v_i-v_j) - N_f \sum_{i,j=1}^{N}
  f(mL,v_i-v_j+\pi) \right] \comma
\end{flalign}
where $m$ is the mass of the fermions and the function $f$ is defined
in terms of the modified Bessel function $K_2$:
\begin{equation}
f(x,v)= \frac{1}{\pi^2} \sum_{g=1}^{\infty} \frac{2 + (gx)^2 K_2(gx)
\cos (gv)}{g^4} \period
\end{equation}

In the large--$N$ limit, the function $v(i/N) = v_i$ can be defined. It
is related to the eigenvalue density $\rho(v(x))=[v'(x)]^{-1}$. Thus,
in the large--$N$ limit, the effective potential is a functional of
$v(x)$:
\begin{flalign}
& \frac{V_{\mathrm{Or}}(v)}{N^2} = \frac{1}{L^4} \int_0^1 \left[
f(0,v(x)-v(y)) - N_f f(mL,v(x)+v(y)+\pi) \right]\ dxdy \comma \\ &
\frac{V_{\mathrm{Adj}}(v)}{N^2} = \frac{1}{L^4} \int_0^1 \left[
f(0,v(x)-v(y)) - N_f f(mL,v(x)-v(y)+\pi) \right]\ dxdy \period
\end{flalign}
In the large--$N$ limit, the center $\Z_N$ acts on the function $v(x)$ as:
\begin{equation}
v(x) \rightarrow v(x) + \alpha \period
\end{equation}
Clearly, the effective potential for the AdjQCD is invariant under the
full center action. Instead, the effective potential for OrientiQCD
transform as:
\begin{equation}
\frac{V_{\mathrm{Or}}(v)}{N^2} \rightarrow \frac{1}{L^4} \int_0^1
\left[ f(0,v(x)-v(y)) - N_f f(mL,v(x)+v(y)+2\alpha+\pi) \right]\ dxdy
\period
\end{equation}
Since $f(x,v)$ has a $2\pi$ period in $v$, the effective potential is
invariant if and only if $\alpha = k\pi$.

This fact has consequences on the vacuum structure of the two
theories. Since in both theories, the terms depending on
$v(x)-v(y)$ generate attraction between the eigenvalues, the minima
are characterised by a $v(x)=\bar{v}$ constant. In the AdjQCD
$\bar{v}$ can take any value, instead in the OrientiQCD the fermionic
term generates only two minima at $\bar{v}=0,\pi$, which correspond to
$P=\pm 1$.

Let us now discuss the implications for the orientifold planar
equivalence. The charge conjugation symmetry acts on the gauge
fields as:
\begin{equation}
A_\mu(x_0,\mathbf{x}) \rightarrow - A_\mu^T(x_0,\mathbf{x}) \period
\end{equation}
The vacua of the OrientiQCD are invariant under charge
conjugation. Instead, only two of the infinite vacua of AdjQCD are
invariant under charge conjugation. These are precisely the two
corresponding to $P=\pm 1$, and the orientifold planar equivalent must
be valid in this two vacua. It can be objected that, since all the
vacua of AdjQCD are unitarily equivalent, the orientifold planar
equivalence should be valid in all the vacua. However the matching
between the observables in the two theories is more involved. Consider
for instance a Wilson loop $W(\Gamma) = \tr U(\Gamma) /N$ along the
closed path $\Gamma$, wrapping $w$ times around the compact
dimension. Let $\langle W(\Gamma) \rangle_{\alpha,\mathrm{Adj}}$ be
its expectation value with respect to the vacua of AdjQCD identified
by $\bar{v}=\alpha$, and let $\langle W(\Gamma) \rangle_{\mathrm{Or}}$
be its expectation value with respect to the vacua of OrientiQCD
identified by $\bar{v}=0$. The following equalities hold:
\begin{equation}
\langle W(\Gamma) \rangle_{\alpha,\mathrm{Adj}} = e^{i\alpha w} \langle W(\Gamma) \rangle_{0,\mathrm{Adj}} = e^{i\alpha w} \langle W(\Gamma) \rangle_{\mathrm{Or}} \comma
\end{equation}
in the planar limit. Therefore the observable $W(\Gamma)$ in the
$\alpha$--vacuum of AdjQCD correspond to the observable $e^{i\alpha w}
W(\Gamma)$ in the $0$--vacuum of OrientiQCD. In the $\alpha$--vacuum of
AdjQCD an unbroken charge conjugation can be defined, by properly
composing the naive one with the center symmetry. On the gauge field,
it acts like:
\begin{subequations}
\begin{flalign}
& A_0(x_0,\mathbf{x}) \rightarrow - A_0^T(x_0,\mathbf{x}) + \frac{2 \alpha}{L} \comma \\
& A_k(x_0,\mathbf{x}) \rightarrow - A_k^T(x_0,\mathbf{x}) \period
\end{flalign}
\end{subequations}
It acts on the Polyakov loop like:
\begin{flalign}
& P(\mathbf{x}) = \tr \mathrm{Pexp} \left[ i \int_0^L A_0(x_0,,\mathbf{x})\ dx_0 \right] \rightarrow \nonumber \\
& \rightarrow e^{2i\alpha} \tr \mathrm{Pexp} \left[ i \int_0^L A_0(x_0,,\mathbf{x})\ dx_0 \right]^* = e^{2i\alpha} P(\mathbf{x})^* \comma
\end{flalign}
and in particular it does not change the expectation value of the Polyakov loop.

\section{The center symmetry in the Hamiltonian formalism}
\label{sec:hamiltonian}

It is instructive to reproduce the results of the previous sections
using the Hamiltonian and coherent states formalism. This approach uncovers a
deeper picture of the center symmetry, and its remnants in the large--$N$ limit.

The large--$N$ limit of gauge theories via coherent states was introduced
in~\cite{Yaffe:1981vf}. This formalism was used in~\cite{Unsal:2006pj} to
prove orientifold planar equivalence, by defining an orientifold projection on
the parent $\mathrm{SO}(N)$ gauge theory with fermions in the adjoint
representation. A review of these concepts is beyond the aims of this paper,
therefore the reader should refer to the bibliography for details.

The center symmetry is not a symmetry of the Hamiltonian. Indeed the
center acts only on the temporal component of the gauge field, which
is not a real degree of freedom of the theory: it is the Langrange
multiplier for the Gauss constraint. Let us review some of the details
here. 

If $H$ is the Hamiltonian of the gauge theory on the lattice,
the partition function is given by:
\begin{equation}
Z = \Tr ( e^{-\beta H} \mathbb{P} ) \comma
\end{equation}
where $\mathbb{P}$ is the projector on the gauge--invariant states. If
$g$ is an element of the gauge group, let
$\mathcal{G}_{\mathbf{x}}[g]$ denote the unitary operator, which acts
on the Hilbert space by producing the gauge transformation $g$ in the
point $\mathbf{x}$. As a consequence of the invariance of the Haar
measure, the projector $\mathbb{P}$ and the partition function can be
written as:
\begin{flalign}
& \mathbb{P} = \int \prod_{\mathbf{x}} \left\{
\mathcal{G}_{\mathbf{x}}[\Omega(\mathbf{x})] d\Omega(\mathbf{x})
\right\} \comma \\ & Z = \int \Tr \left( e^{-\beta H}
\prod_{\mathbf{x}} \mathcal{G}_{\mathbf{x}}[\Omega(\mathbf{x})]
\right) \prod_{\mathbf{x}} d\Omega(\mathbf{x}) \period
\end{flalign}
If $\{ \psi_n \}$ is a basis for the Hilbert space, the trace in the
integral can be written as:
\begin{equation}
\Tr \left( e^{-\beta H} \prod_{\mathbf{x}}
\mathcal{G}_{\mathbf{x}}[\Omega(\mathbf{x})] \right) = \sum_n \langle
\psi_n | e^{-\beta H} | \psi_n^{(\Omega)} \rangle \comma
\end{equation}
where $\psi_n^{(\Omega)}$ is obtained by applying the gauge
transformation $\Omega$ to the state $\psi_n$. From the equation
above, $\Omega(\mathbf{x})$ is the $\mathrm{SU}(N)$ phase that the state
$\psi_n$ acquires after a translation around the temporal
direction. By writing the matrix element of $e^{-\beta H}$ as a
functional integral, one sees that $\Omega(\mathbf{x})$ is the
parallel transport around the time direction and $ \tr
\Omega(\mathbf{x}) $ is the Polyakov loop.

All the potentials of Sect.~\ref{sec:effpotential} can be
written using the Hamiltonian formalism. For instance, the probability
distribution of the parallel transport $\Omega(\bfx)$ is:
\begin{equation}
e^{ -S_\Omega(\Omega) } = \frac{1}{Z} \Tr 
\left\{ e^{-\beta H} \prod_{\mathbf{x}} \mathcal{G}_{\mathbf{x}}
[\Omega(\mathbf{x})] \right\} \comma
\end{equation}

If $u$ is an element of the center, a center transformation is defined
as $\Omega(\bfx) \rightarrow u \Omega(\bfx)$. It does not
affect the degrees of freedom in the Hamiltonian, but only the Gauss
constraint:
\begin{equation}
e^{ -S_\Omega(u\Omega) } = \frac{1}{Z} \Tr \left\{ e^{-\beta H}
\prod_{\mathbf{x}} \mathcal{G}_{\mathbf{x}}[u\Omega(\bfx)] \right\}
\period
\end{equation}
When a center symmetry is present, it is not a symmetry for the
quantum system in a proper sense: it is not implemented by a unitary
operator on the Hilbert space. It is a symmetry only of the potential
$S_\Omega(\Omega)$.

This fact implies that an analysis of the center symmetry in the
large--$N$ limit using the coherent states formalism cannot be
developed in a straightforward way. Indeed, in the coherent state
formalism, the Gauss constraint is completely solved by taking only
gauge--invariant observables as degrees of freedom. Thus, no analog of
the Polyakov loop exists.

This problem can be circumvented as follows.  Rotate the system (in
Euclidean space--time) and interpret the gauge theory on an infinite
space and at finite temperature, as the same theory at zero
temperature and on a space $S_1 \times \mathbf{R}^2$ with antiperiodic
boundary conditions for the fermions. After the rotation, the Polyakov
loop becomes the parallel transport around the spatial compact
dimension, and the center acts now on the physical degrees of freedom
of the theory. If $\Sigma$ is a plane orthogonal to the compact
dimension, and $\plinks(\Sigma)$ is the set of all the positive links
departing from sites of $\Sigma$ and orthogonal to it, the center acts
as:
\begin{equation}
U_\ell \rightarrow u U_\ell \quad \textrm{if } \ell \in
\plinks(\Sigma) \period
\end{equation}
Consider a Wilson loop $W(\Gamma)$ in the spatial lattice. The center
symmetry counts the winding number $w(\Gamma)$ of the Wilson loop:
\begin{equation}
W(\Gamma) \rightarrow u^{w(\Gamma)} W(\Gamma) \period
\end{equation}
It is clear that, unless $u \in \Z_2$, the center mixes the real and
imaginary parts of the Wilson loops, therefore it does not commute
with the charge conjugation symmetry.

Let us come now to the orientifold planar equivalence. The $SO(N)$
parent theory has a $\Z_2$ symmetry in the large--$N$ limit. This
symmetry is mapped through the orientifold projection in the $\Z_2$
symmetry of the OrientiQCD in the large--$N$ limit. Instead, the
AdjQCD has a $\Z_N$ symmetry, but only the $\Z_2$ subgroup in the large--$N$
limit maps the neutral sector into itself. We
conclude once more, that $\Z_2$ (and not $\Z_N$) is the only symmetry
of OrientiQCD that we can deduce from the orientifold planar
equivalence.

\section{Conclusions}

In this work we addressed the issue of the center symmetry of
orientifold theories, by considering $\mathrm{SU}(N)$ gauge theories with
fermions in the symmetric/antisymmetric two--index representations in
the large--$N$ limit. Our approach is based on the idea that the right
tool for an exhaustive analysis of the center symmetry is the quantum
action (or equivalently the generator of the connected correlators)
for the Polyakov loop in the theory at finite temperature. We
investigated the quantum action in both the confined and the
deconfined phases and we conclude that the orientifold theory is
invariant under the $\Z_2$ subgroup of the center.

In the deconfined phase the invariance under (the spontaneously broken) $\Z_2$ implies the
well--known existence of only two degenerate vacua. The analytic
computation can be carried out in the  high--temperature regime,
where the one--loop approximation holds. In this case
the effective potential of the Polyakov loop has two degenerate minima
corresponding to $\langle P \rangle/N = \pm 1$.

In the confined phase the generator of the connected correlators of
Polyakov loops can be analytically computed in the large--mass
expansion of the lattice theory. We showed that all the terms, that
are not invariant under $\Z_2$, vanish in the large--$N$ limit. On the
other hand the $\Z_2$--invariant terms are different from zero,
showing that actually the center symmetry group is explicitly broken to
the $\Z_2$ subgroup in the large--$N$ limit.

A previous analysis of the center symmetry of the orientifold theories
can be found in Ref.~\cite{Armoni:2007kd}, where the authors concluded
that the $\Z_N$ center symmetry is dynamically recovered in the
large--$N$ limit at zero temperature: the part of the fermionic action
that is not invariant under the action of $\Z_N$ decouples from the
expectation values in the large--$N$ limit in the confined phase. We
want to point out that this result was obtained for observables with
trivial topology with respect to the temporal compact dimension. Only
these observables are relevant at zero temperature. In particular the
dynamics of Wilson loops is determined by the $\Z_N$ invariant part of
the fermionic action and this fact implies the stability of the
$k$--strings. However at non--zero temperature, the theory contains
also observables with non--trivial topology with respect to the
temporal compact dimension. We showed explicitly that the Polyakov
loop couples to the part of the fermionic action with non--trivial
winding number and this is in general a leading effect in the
large--$N$ limit. Therefore, although the $\Z_N$ symmetry is
dynamically recovered in a subsector, the whole orientifold theory is
invariant only under the action of $\Z_2$.

This picture fits well with the predictions of the orientifold planar
equivalence. In the large--$N$ limit the orientifold theory is
equivalent to an $\mathrm{SU}(N)$ theory with Majorana fermions in the
adjoint representation in a neutral sector, defined by all the
single--trace gauge--invariant C--even observables. The latter theory
is invariant under the full $\Z_N$ center symmetry. Although
the two theories have different symmetry contents, we showed that this
fact is not in contradiction with the orientifold planar
equivalence. Indeed the equivalence is valid only between neutral
sectors, thus we can expect the same symmetry in the two theories only
after removing all the states and the observables outside of these
neutral sectors. In other word, the orientifold planar equivalence
implies the equality in the two theories of the symmetry subgroups
that map the neutral sector into itself. In the case of the center
symmetry, this subgroup is $\Z_2$ for both theories.

\bigskip
{\bf Acknowledgements} LDD is supported by an STFC Advanced Fellowship. AP by an
STFC SPG grant. The authors wish to thank Adi Armoni, Mithat Unsal and Gabriele
Veneziano for useful discussions and comments.

\appendix

\section{Lattice theory and large--$N$ counting}
\label{app:a}

In this work we are interested in computing connected expectation
values of products of closed loops in pure Yang--Mills theories.  The
Boltzmann weight $e^{-S}$ in the path integral can be expanded in a
series in $1/\lambda$. The expansion of the Wilson gauge action
produces a series of monomials of elementary plaquettes. The
expectation values are computed by integrating each link variable over
the group manifold, and non--vanishing results are obtained only when
the plaquettes from the expansion of the action produce a tiling of a
surface whose boundary is given by the closed loops. Each plaquette in
this expansion appears with a factor of $N$. The rules for the
$\mathrm{SU}(N)$ group integration are known in detail, and the result
of integrating over the group manifold a generic product of matrix
elements can be found e.g. in
Refs.~\cite{Bars:1979xb,GonzalezArroyo:1999bc}. The integration
contributes further factors of $1/N$:
\begin{flalign}
\int dU U_{i_1 j_1} U^\dagger_{l_1 m_1} &=
\frac1N \delta_{i_1,m_1} \delta_{j_1,l_1}\comma \label{eq:app:twolinks} \\
\int dU U_{i_1 j_1} U_{i_2 j_2} U^\dagger_{l_1 m_1} U^\dagger_{l_2
  m_2} &=
\frac{1}{N^2-1} \left(
\delta_{i_1,m_1} \delta_{j_1,l_1} \delta_{i_2,m_2} \delta_{j_2,l_2} +
\delta_{i_1,m_2} \delta_{j_1,l_2} \delta_{i_2,m_1} \delta_{j_2,l_1} \right)
\nonumber \\
& -
\frac{1}{N(N^2-1)} \left(
\delta_{i_1,m_1} \delta_{j_2,l_2} \delta_{i_2,m_2} \delta_{j_1,l_1} +
\delta_{i_1,m_2} \delta_{j_2,l_1} \delta_{i_2,m_1} \delta_{j_1,l_2} 
\right) \comma
\nonumber \\
& \label{eq:app:fourlinks} \\
\int dU U_{i_1 j_1} \cdots U_{i_N j_N} &= \frac{1}{N!} \epsilon_{i_1 \dots i_N} \epsilon_{j_1 \dots j_N}
\period \label{eq:app:Nlinks}
\end{flalign}
Each trace over colour indices $(i,j,l,m)$ contributes a factor of
$N$. Collecting all contributions one obtains that each diagram is
proportional to $N^\chi$, where $\chi$ is the Euler characteristics of
the surface spanned by the tiling, see
e.g. Ref.~\cite{tHooft:2002yn}. 

For the diagrams considered in this work:
\begin{equation}
\chi=2-2H-2B\comma
\end{equation}
where $H$ is the number of handles, and $B$ the number of boundaries
of the surface. The number of boundaries is given by the number of
closed loops. One can readily see that the planar limit is obtained by
considering surfaces with $H=0$.

Let us now describe in detail the computations that appear in the
derivation of the results in Sect.~\ref{sec:effpotential}.

Consider the connected expectation values:
\begin{equation} 
\label{eq:app:connected}
\langle \prod_{i=1}^n \tr U(\omega_i) \rangle_{c,\mathrm{YM}} \period
\end{equation}
We will prove that the leading $N^{2-n}$ behaviour is saturated if
loops with winding number different from zero only appear in pairs
with opposite winding numbers.

The easiest case is $\langle \tr U(\omega_1) \tr U(\omega_2)
\rangle_{\mathrm YM,c}$, with $w_1 = -w_2$. In this case the surface
connecting the two loops is a cylinder, wrapping $|w_1|$ times around
the thermal dimension (see Fig.~\ref{fig:ym1}). Notice that it is
essential that the loops have opposite directions, since the cylinder
cannot be twisted because of the topology of the torus. Since the
surface is planar, this diagram saturates the leading behaviour of the
connected expectation value:
\begin{equation}
\langle \tr U(\omega_1) \tr U(\omega_2) \rangle_{\mathrm YM,c} = O(N^0) \qquad \textrm{with } w_1 = -w_2 \period
\end{equation}

Consider now the more complex case of $\langle \tr U(\omega_1) \tr
U(\omega_2) \tr U(\omega_3) \tr U(\omega_4) \rangle_{\mathrm YM,c}$
with $w_1=-w_2$ and $w_3=-w_4$. The planar surface is built by
connecting each pair of loops with a cylinder with an hole, and than
gluing the two cylinders through the boundary of the holes (see
Fig.~\ref{fig:ym1}). The Euler characteristic of this surface is $\chi
= -2$ therefore the corresponding diagram behaves like $N^{-2}$.

The above procedure can be iteratively generalised to the case of an
arbitrary number of pairs of loops. Moreover an arbitrary number of
loops with zero winding number can be included simply adding holes to
the surface. The resulting surface is always planar.

Consider again the generic connected expectation value in
Eq.~(\ref{eq:app:connected}). If some loop $\omega_i$ with winding
number $w_i$ different from zero is not paired to a loop with winding
number $-w_i$ then the leading behaviour cannot be saturated. Since
this fact is not crucial for our work, we only illustrate it in
some simple cases.

For $n=3$ and $w_1=w_2=1$, $w_3=-2$, the connected expectation value
$$\langle \tr U(\omega_1) \tr U(\omega_2) \tr U(\omega_3)
\rangle_{\mathrm YM,c}$$
has zero net winding number. However it is not
possible to build a surface connecting this three loops because of the
topology of the torus. One can ask if it is possible to use the
four--link integration to get a leading contribution. We shall proceed
in steps to prove that this is not the case. 

First of all, we can deform each loop by taking a contiguous plaquette
from the expansion of the Bolzmann weight and integrating the common
link, as depicted in Fig.~\ref{fig:deformation}. If $\tr (AU)$ is
schematically the loop and $N \lambda^{-1} \tr (U^\dagger B)$ is the
relevant term of the action:
\begin{equation} \label{eq:app:deformation}
N \lambda^{-1} \int dU \tr(AU) \tr(U^\dagger B) = \lambda^{-1} \tr
(AB) \period
\end{equation}
In this way, the loop is replaced by the deformed one times a factor
$1/\lambda$. We can use iteratively this procedure for each loop until
the glued plaquettes shape three cylinders that end in three loops in
the same spatial point (Fig.~\ref{fig:threeloops}). Now we start to
integrate one of the links of these three loops, by using the
four--link formula in Eq.~(\ref{eq:app:fourlinks}). If we want to
integrate the link $U$, we can write the three loops schematically as
$\tr(A U)$, $\tr(A U)$, $\tr(U^\dagger A^\dagger U^\dagger A^\dagger)$
(remember that they all are coincident). The result of the integration
is:
\begin{flalign}
& \int dU \tr(A U) \tr(A U) \tr(U^\dagger A^\dagger U^\dagger
A^\dagger) = \nonumber \\ & \qquad = \frac{2}{N^2-1} \tr(A A^\dagger A
A^\dagger) - \frac{2}{N(N^2-1)} [\tr(A A^\dagger)]^2 = \nonumber \\ &
\qquad = \frac{2 N}{N^2-1} - \frac{2 N^2}{N(N^2-1)} = 0 \period
\label{eq:app:threeloops}
\end{flalign}
In the last line we used the fact that $A$ is the product of all the
links around the loop but $U$, and therefore it is a unitary
matrix. We conclude that this kind of diagrams does not contribute to
the connected expectation value.

%\begin{figure}[!ht]
\FIGURE{
  \centering
  \epsfig{file=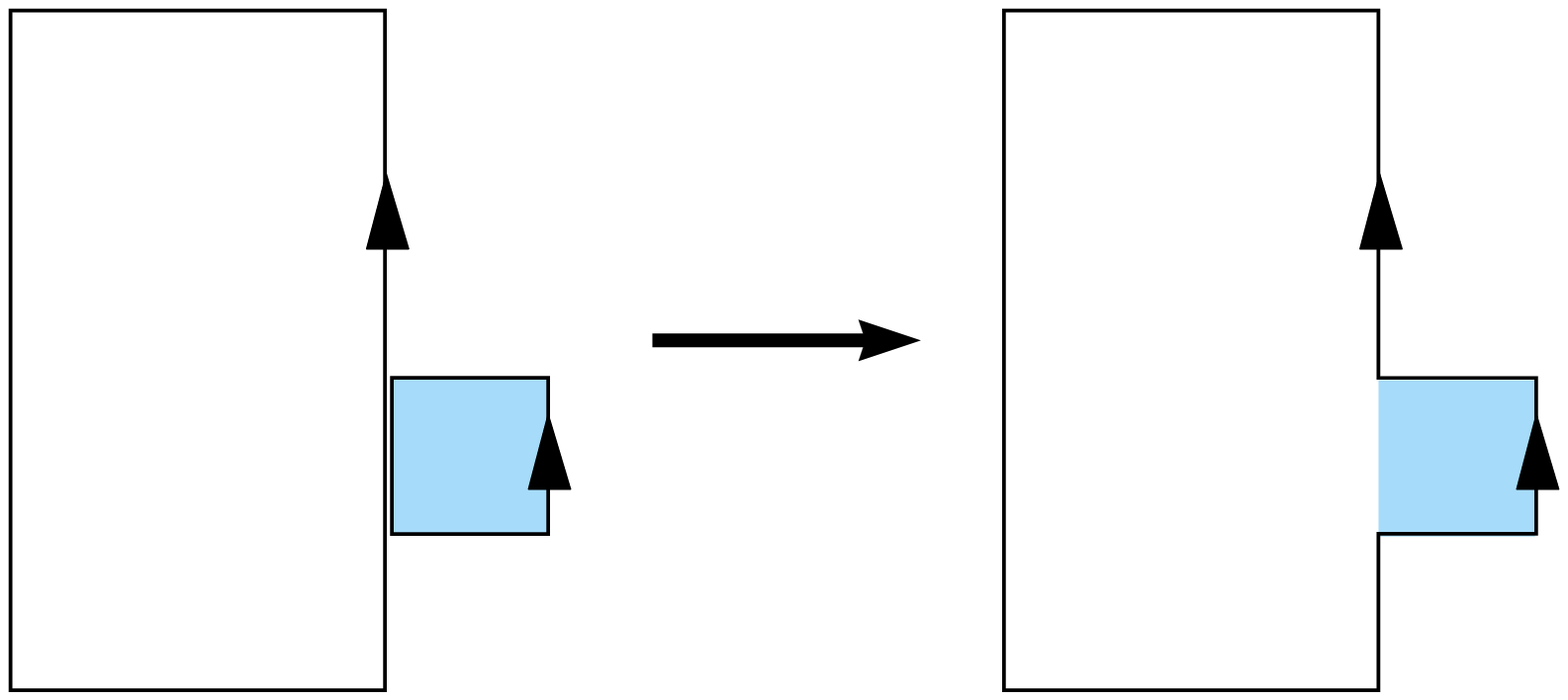,scale=0.6}  
  \caption{The Eq.~(\ref{eq:app:deformation}) is represented. In the
  left side, the big loop $\tr(AU)$ and the plaquette $\tr(U^\dagger
  B)$; in the right side, the deformed loop $\tr (AB)$.}
  \label{fig:deformation}
}
%\end{figure}

%\begin{figure}[!ht]
\FIGURE{
  \centering
  \epsfig{file=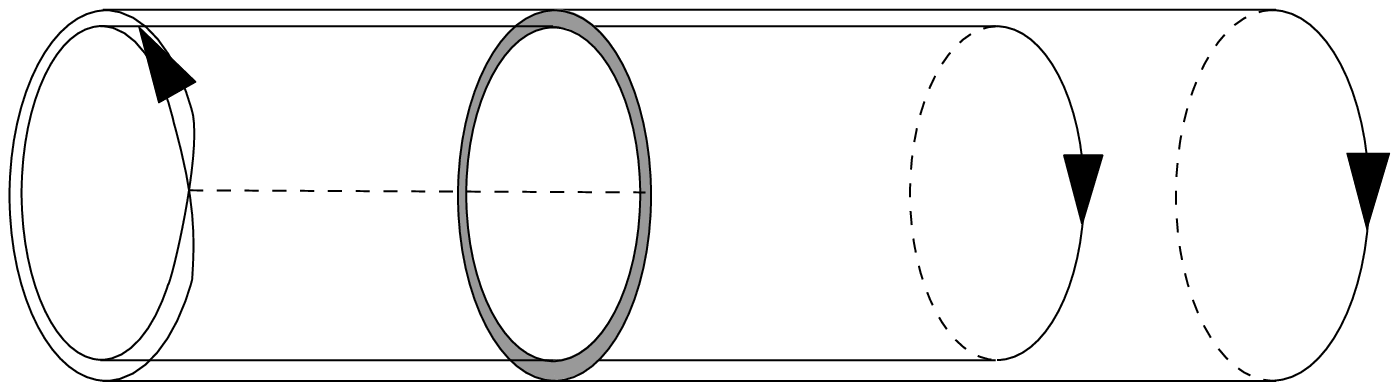,scale=0.8}  
  \caption{The connected expectation value $\langle \tr U(\omega_1)
  \tr U(\omega_2) \tr U(\omega_3) \rangle_{\mathrm YM,c}$ with
  $w_1=w_2=1$, $w_3=-2$. The four--link integration in the gray area
  makes vanish this contribution, as explained in
  Eq.~(\ref{eq:app:threeloops}).}
  \label{fig:threeloops}
}
%\end{figure}

The last non--trivial case we want to illustrate is $n=1$ and $w_1=N$.
In this case the connected expectation value $\langle \tr U(\omega_1)
\rangle_{\mathrm YM,c}$ is invariant under the center $Z_N$, but not
$\mathrm{U}(1)$. Therefore a non--zero contribution can be constructed only
using the $N$--link integration formula for $\mathrm{SU}(N)$ in
Eq.~(\ref{eq:app:Nlinks}). By using the same construction as above,
the loop $\tr U(\omega_1)$ can be deformed (without introducing extra
$N$ factors) into a loop wrapping straight in the thermal direction
(see Fig.~\ref{fig:ym2}), that can be written schematically as $\tr
[(AU)^N]$. Integrating the link $U$:
\begin{flalign}
\int dU \tr[(UA)^N] & = \frac{1}{N!} \epsilon_{i_1 \dots i_N} A_{i_1 j_1} \cdots A_{i_N j_N} \epsilon_{j_1 \dots j_N} = \nonumber \\
& = \det A = 1 \period
\end{flalign}
This is a subleading contribution since the leading term of the
expectation value of a single loop is expected to be proportional to
$N$.

%\bibliography{bib/hirep,bib/alg,bib/QFT,bib/lat,bib/largeN}

\bibliography{paper}

\end{document}